\documentclass[12pt]{iopart}
\usepackage{amssymb}
\usepackage{epsf}

\begin{document}

\title[Scale-invariant universal crossing probability]{Scale-invariant 
universal crossing probability in one-dimensional diffusion-limited 
coalescence}

\author[L. Turban]{L. Turban}

\address{Laboratoire de Physique des Mat\'eriaux, Universit\'e Henri 
Poincar\'e (Nancy 1), BP~239, 
F-54506 Vand\oe uvre l\`es Nancy Cedex, France}

\ead{turban@lpm.u-nancy.fr}

\begin{abstract}
The crossing probability in the time direction, $\pi_t$, is defined for an 
off-equilibrium reaction-diffusion system as the probability that the system
of size $L$ is still active at time $t$, in the finite-size 
scaling limit. Exact results are obtained for the diffusion-limited 
coalescence problem in $1+1$ dimensions with periodic and free boundary 
conditions using empty interval methods. $\pi_t$ is a scale-invariant 
universal function of an effective aspect ratio, $L^2/Dt$, which is the 
natural scaling variable for this strongly anisotropic system.  
\end{abstract}

%\submitto{\JPA}

%\pacs{05.40.-a, 05.70.Ln, 82.20.-w}

%\maketitle % for title page

\section{Introduction}
The study of crossing probabilities for standard percolation has been the 
subject of much interest during the last decade~[1--16] (for a recent 
introductory review see reference~\cite{cardy01}). 

In two dimensions, the crossing probability may be defined as the 
probability $\pi$ to have at least one cluster joining two opposite edges of 
a rectangular-shaped finite system with length~$L_\parallel$ and 
width~$L_\perp$. It turns out that, at the percolation threshold, in the 
finite-size scaling limit ($L_\parallel\to\infty$, $L_\perp\to\infty$, with 
$r=L_\perp/L_\parallel$ fixed), the crossing probability is a {\it 
scale-invariant universal function}, $\pi(r)$, of the aspect ratio 
$r$~\cite{langlands92,cardy92}.   

Following the numerical work of Langlands {\it et al}~\cite{langlands92}, 
Cardy~\cite{cardy92} was able to derive an exact expression for $\pi(r)$. 
Using the relation between percolation and the $q$-state Potts model in the 
limit $q\to1 $~\cite{kasteleyn72} and boundary conformal field theory, he 
obtained the crossing probability between two non-overlapping segments on 
the edge of the half-plane at criticality. The corresponding result in the 
rectangular geometry was then obtained through a conformal mapping. The 
non-trivial {\it scale invariance} of $\pi(r)$ is linked to the vanishing of 
the scaling dimension $x(q)$ of a boundary condition changing operator of the 
Potts model in the percolation limit, $q\to 1$. Let us notice that some of 
these results have been rigorously proven 
recently~\cite{smirnov01,schramm01}. A related problem concerns the number of 
incipient spanning clusters at criticality~[12--16]. Exact formulas have been 
also obtained in this field through conformal and Coulomb-gas 
methods~\cite{cardy98,cardy02}.

In a recent work~\cite{turban02} the critical crossing probability was 
studied numerically for a strongly anisotropic system, namely, directed 
percolation in $1+1$ dimensions. In this case, the crossing probability in 
the time direction $\pi_t$ is also the probability that the system of 
size~$L$ remains active at time~$t$. Anisotropic 
scaling~\cite{binder89,hucht02} then implies that the appropriate aspect 
ratio is $r=L^z/t$, where $z$ is the dynamical exponent. Here, too, it was 
found that, in the finite-size scaling limit, the critical crossing 
probability is a scale-invariant universal function of an effective 
aspect ratio which is the product of $r$ by a non-universal constant.

In the present work we continue the examination of critical crossing 
probabilities in strongly anisotropic systems by considering the case of 
diffusion-limited coalescence (DLC). This is one of the many actively studied 
off-equilibrium systems~[22--26] which yields itself to an exact 
analysis~[27--40]. We study the problem with periodic and free boundary 
conditions using the empty interval method, or interparticle distribution 
function method, which is reviewed in~\cite{benavraham95,benavraham97}. 

The case of periodic boundary conditions is studied in section~2 using the 
standard empty interval method for a finite discrete system~[33--35]. In 
section~3 we use a modification of the standard method to treat the problem 
with free boundary conditions. The results are discussed in section~3.

\section{Diffusion-limited coalescence with periodic boundary conditions}

We consider the time evolution of DLC on a one-dimensional lattice with $L$ 
sites and periodic boundary conditions. Each site is in one of two states, 
either vacant or occupied by a particle $A$. The dynamics is governed by the 
following processes
\begin{equation}
A\,\varnothing\ \stackrel{D}{\longleftrightarrow}\ \varnothing\,A\quad
{\rm(diffusion)}\,,
\qquad
A\,A\ \stackrel{D}{\longrightarrow}\ \cases{A\,\varnothing\\ \varnothing\,A}
\quad{\rm(coagulation)}\,,
\label{e2.proc}
\end{equation}
with the {\it same} rate $D$. When a particle jumps with rate $D$ on a 
nearest-neighbour site which is already occupied, the two particles 
coalesce {\it immediately} on this site. Thus coagulation may occur either 
to the left or to the right. To simplify we assume that the $L$ sites are 
occupied with probability one in the initial state at $t=0$. As a 
consequence, the probability distribution of the particles $A$ is 
translation invariant at later time $t\geq0$. 

We study the time evolution of the system using the empty interval  
method~\cite{benavraham90}. Let the symbol $\bullet$ ($\circ$) denote an 
occupied (vacant) site. The probability for a given interval of length $n$ to 
be empty at time $t$,
\begin{equation}
I_n(t)={\rm Prob}\,\big(\overbrace{\circ\circ\cdots\circ\circ}^{n}\big)\,,
\label{e2.in}
\end{equation}
is translation invariant on the periodic system with uniform initial 
conditions. Its time evolution involves the probability 
\begin{equation}
F_n(t)
={\rm Prob}\,\big(\bullet\overbrace{\circ\circ\cdots\circ\circ}^{n}\big)
={\rm Prob}\,\big(\overbrace{\circ\circ\cdots\circ\circ}^{n}\bullet\big)
\label{e2.fn}
\end{equation}
to have an empty interval of length $n$, either preceded or followed by an 
occupied site. Since
\begin{equation}
{\rm Prob}\,\big(\overbrace{\circ\circ\cdots\circ\circ}^{n}\bullet\big)
+{\rm Prob}\,\big(\overbrace{\circ\circ\cdots\circ\circ\,\circ}^{n+1}\big)
={\rm Prob}\,\big(\overbrace{\circ\circ\cdots\circ\circ}^{n}\big)\,,
\label{e2.fisymb}
\end{equation}
the following relation is obtained:
\begin{equation}
F_n(t)=I_n(t)-I_{n+1}(t)\,.
\label{e2.fnin}
\end{equation}
For $n=1,L-1$, the empty interval probability satisfies the master equation
\begin{equation}
{{\rm d} I_n(t)\over{\rm d} t}=2D\,[F_{n-1}(t)-F_n(t)]
=2D\,[I_{n-1}(t)-2I_n(t)+I_{n+1}(t)]\,.
\label{e2.mast}
\end{equation}
The gain terms correspond to processes in which a particle occupying the 
first site on the right (left) of an empty interval of length $n-1$ jumps 
to the right (left) to diffuse or coalesce on the next site, thus leaving 
behind an empty interval of length $n$. The loss terms correspond to 
processes in which a nearby particle enters an empty interval of length 
$n$ either from the left or from the right. The final form of the 
difference equation follows from~\eref{e2.fnin}.

Equation~\eref{e2.mast} has to be solved with the boundary conditions
\begin{equation}
I_0(t)=1\,,\qquad I_L(t)=0\,.
\label{e2.bc}
\end{equation}
The first one results from the expression of the site occupation 
probability, $F_0(t)=1-I_1(t)$, the second follows from the fact that an 
initially non-empty system remains so at later time, since the 
coalescence process leaves at least one surviving particle. With a full 
lattice at $t=0$, the initial condition corresponds to 
\begin{equation}
I_n(0)=\delta_{n,0}\,.  
\label{e2.ic}
\end{equation}

The master equation~\eref{e2.mast} is solved through the Ansatz
\begin{equation}
I_n(t)=\sum_q\phi_q(n)\,\e^{-\omega_q t}
\label{e2.ansatz}
\end{equation}
where $\phi_q(n)=u_q\sin(qn)+v_q\cos(qn)$ when $\omega_q=8D\sin^2(q/2)$ is 
non-vanishing. It takes the form $\phi_0(n)=an+b$ for the zero mode,  
$\omega_0=0$, corresponding to the stationary state. The first boundary 
condition in~\eref{e2.bc}, 
\begin{equation}
I_0(t)=1=b+\sum_{q\neq0}v_q\,\e^{-\omega_q t}\,,
\label{e2.bc1}
\end{equation}
leads to $b=1$ and $v_q=0$ whereas the second,
\begin{equation}
I_L(t)=0=1+aL+\sum_{q\neq0}u_q\sin(qL)\,\e^{-\omega_q t}\,,
\label{e2.bc2}
\end{equation}
gives $a=-1/L$ and $\sin(qL)=0$. Thus the empty interval probability can 
be written as~\cite{henkel01}
\begin{equation}
I_n(t)=1-{n\over L}+\sum_{k=1}^{L-1}c_k\sin\!\left({nk\pi\over L}\right)
\exp\!\left[-8Dt\sin^2\!\left({k\pi\over 2L}\right)\right]\,.
\label{e2.solgen}
\end{equation}
The stationary state solution $I_n(\infty)=1-n/L$ corresponds to a single 
particle diffusing on the $L$ sites so that, a site being occupied with 
probability $1/L$, an interval of $n$ sites is non-empty with probability 
with probability $n/L$.

According to~\eref{e2.ic}, at $t=0$ we have:
\begin{equation}
\sum_{k=1}^{L-1}c_k\sin\!\left({nk\pi\over L}\right)={n\over L}-1\,,
\qquad n\neq0\,.
\label{e2.ic1}
\end{equation}
Making use of the orthogonality relation for the sines,
\begin{equation}
\sum_{n=1}^{L-1}\sin\!\left({nk\pi\over L}\right)
\sin\!\left({nl\pi\over L}\right)
={L\over2}\,\delta_{k,l}\,,\qquad (k,l=1,L-1)\,,
\label{e2.ortho}
\end{equation}
in equation~\eref{e2.ic1}, we obtain
\begin{equation}
{L\over2}c_k={S_1(k)\over L}-S_2(k)\,,
\label{e2.al}
\end{equation}
where
\begin{eqnarray}
S_1(k)&=\sum_{n=1}^{L-1}n\sin\!\left({nk\pi\over L}\right)
=(-1)^{k+1}{L\over2}\cot\!\left({k\pi\over2L}\right)\,,\nonumber\\
S_2(k)&=\sum_{n=1}^{L-1}\sin\!\left({nk\pi\over L}\right)
={1-(-1)^k\over2}\cot\!\left({k\pi\over2L}\right)\,,
\label{e2.sum}
\end{eqnarray}
so that finally
\begin{equation}\fl
I_n(t)=1-{n\over L}-{1\over L}
\sum_{k=1}^{L-1}\cot\!\left({k\pi\over2L}\right)
\sin\!\left({nk\pi\over L}\right)
\exp\!\left[-8Dt\sin^2\!\left({k\pi\over 2L}\right)\right]\,.
\label{e2.sol}
\end{equation}
The mean number of particles per site (or site occupation probability),
$\rho(t)=1-I_1(t)$ has the well-known $t^{-1/2}$ long-time behaviour in 
the infinite system~\cite{peliti85,spouge88}. Here we are interested in the 
behaviour of the crossing probability in a system with aspect ratio $r=L^z/t$ 
with a dynamical exponent $z=2$ for DLC. The crossing probability in the 
time direction, $P_t(L,t)$, is the probability that the system of size 
$L$ is still active at time~$t$. 

%%%%%%%%%%%%%%%%%%%%%%%%%%%%%%%%%%%%%%%%%%%%%%%%%%%%%%%%%%%%%%%%%%%%
\begin{figure}[tbh]
\vglue3mm
\epsfxsize=7cm
\begin{center}
\mbox{\epsfbox{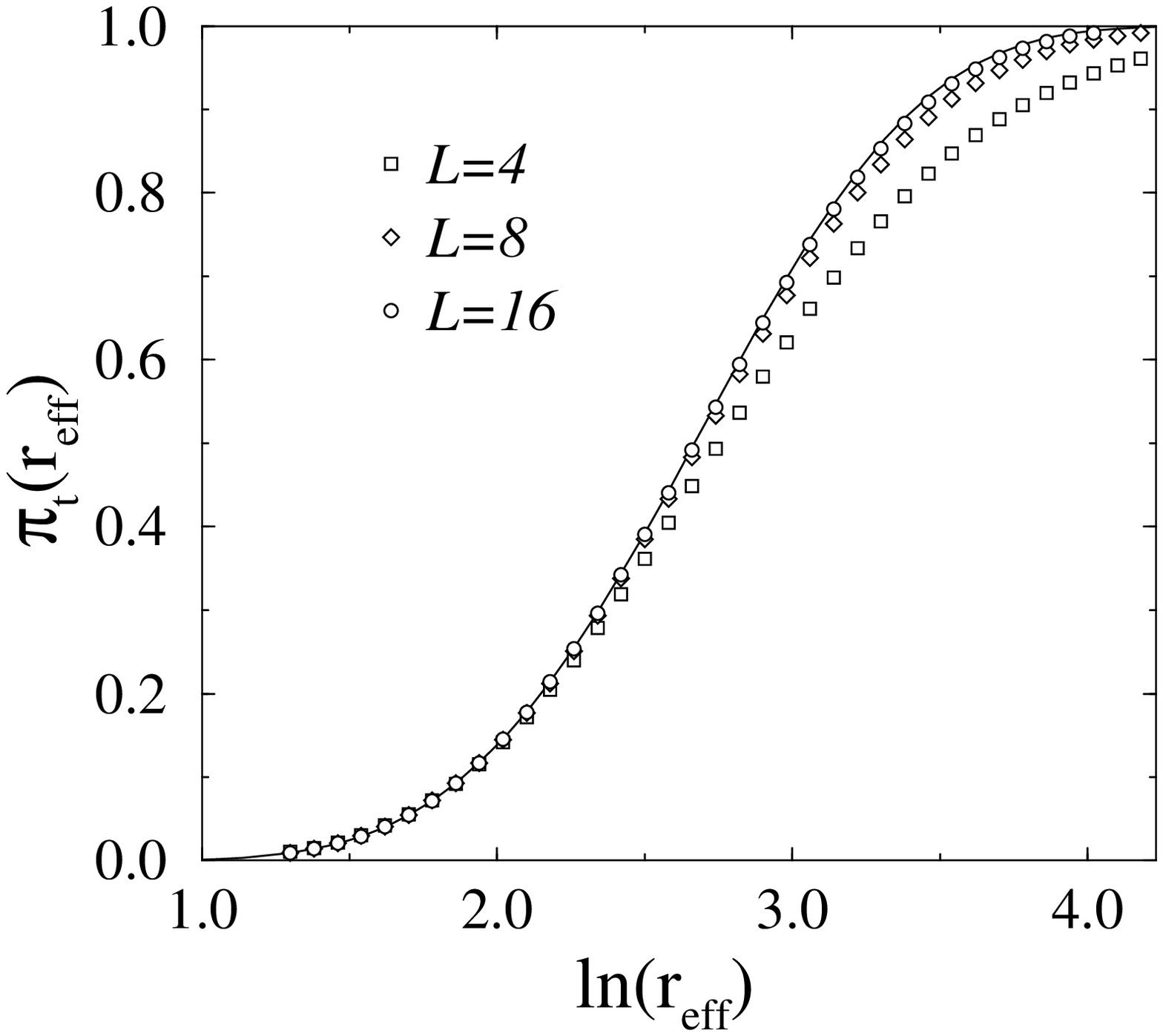}}
\end{center}
\vglue-3mm
\caption{\label{fig1} Variation of the scale-invariant crossing 
probability with the effective aspect ratio for periodic boundary 
conditions. The solid line corresponds to the asymptotic expression in 
equation~\protect\eref{e2.pit}.}
\end{figure}
%%%%%%%%%%%%%%%%%%%%%%%%%%%%%%%%%%%%%%%%%%%%%%%%%%%%%%%%%%%%%%%%%%%%

The probability that the system is in the stationary state, with the last 
particle on a given site, is equal to the probability $I_{L-1}(t)$ that 
the $L-1$ other sites are empty. Since there are $L$ possible choices for 
the occupied site, one obtains:
\begin{equation}\fl
P_t(L,t)=1-LI_{L-1}(t)=2\sum_{k=1}^{L-1}(-1)^{k+1}
\cos^2\!\left({k\pi\over2L}\right)
\exp\!\left[-8Dt\sin^2\!\left({k\pi\over 2L}\right)\right]\,.
\label{e2.pilt}
\end{equation}
In the finite-size scaling limit, this leads to the scale-invariant 
expression
\begin{equation}
\pi_t(r_{\rm eff})=\lim_{\scriptstyle L,t\to\infty\atop
\scriptstyle r\;\mbox{\scriptsize fixed}}P_t(L,t)
=2\sum_{k=1}^\infty(-1)^{k+1}\exp\!\left(-{2k^2\pi^2\over 
r_{\rm eff}}\right)+\Or(L^{-2})\,,
\label{e2.pit}
\end{equation}
where the crossing probability $\pi_t$ shown in figure~\ref{fig1} is a 
universal function of the effective aspect ratio $r_{\rm eff}=r/D=L^2/Dt$.

\section{Diffusion-limited coalescence with free boundary conditons}

\subsection{Master equations for the empty interval probabilities}

With free boundary conditions a modified version of the empty interval 
method is needed to calculate the crossing probability. We define
\begin{equation}
J_{m,n}(t)={\rm Prob}\,
\big(\stackrel{1}{\circ}\circ\cdots\stackrel{m}{\circ}\quad
\stackrel{n+1}{\circ}\!\!\cdots\circ\stackrel{L}{\circ}\big)
\label{e3.jmn}
\end{equation}
as the probability to have two disconnected empty intervals with sites $1$ 
to $m$ and $n+1$ to $L$ empty. Its time evolution depends on the 
probabilities 
\begin{eqnarray}
G_{m,n}(t)&={\rm Prob}\,
\big(\stackrel{1}{\circ}\circ\cdots\circ\!\!\!\stackrel{m+1}{\bullet}\quad
\stackrel{n+1}{\circ}\!\!\!\cdots\circ\stackrel{L}{\circ}\big)\,,
\nonumber\\
H_{m,n}(t)&={\rm Prob}\,
\big(\stackrel{1}{\circ}\circ\cdots\stackrel{m}{\circ}\quad
\stackrel{n}{\bullet}\circ\cdots\circ\stackrel{L}{\circ}\big)\,.
\label{e3.ghmn}
\end{eqnarray}
The different probabilities satisfy the relations
\begin{eqnarray}
\fl
{\rm Prob}\,\big(\stackrel{1}{\circ}\circ\cdots\stackrel{m}{\circ}&\quad
\stackrel{n+1}{\circ}\!\!\cdots\circ\stackrel{L}{\circ}\big)=\nonumber\\
&={\rm Prob}\,
\big(\stackrel{1}{\circ}\circ\cdots\circ\!\!\!\stackrel{m+1}{\bullet}\quad
\stackrel{n+1}{\circ}\!\!\!\cdots\circ\stackrel{L}{\circ}\big)
+{\rm Prob}\,
\big(\stackrel{1}{\circ}\circ\cdots\circ\!\!\!\stackrel{m+1}{\circ}\quad
\stackrel{n+1}{\circ}\!\!\!\cdots\circ\stackrel{L}{\circ}\big)\nonumber\\
&={\rm Prob}\,
\big(\stackrel{1}{\circ}\circ\cdots\stackrel{m}{\circ}\quad
\stackrel{n}{\bullet}\circ\cdots\circ\stackrel{L}{\circ}\big)
+{\rm Prob}\,
\big(\stackrel{1}{\circ}\circ\cdots\stackrel{m}{\circ}\quad
\stackrel{n}{\circ}\circ\cdots\circ\stackrel{L}{\circ}\big)
\label{e3.jgh1}
\end{eqnarray}
so that we have
\begin{eqnarray}
G_{m,n}(t)&=J_{m,n}(t)-J_{m+1,n}(t)\,,\nonumber\\
H_{m,n}(t)&=J_{m,n}(t)-J_{m,n-1}(t)\,.
\label{e3.jgh2}
\end{eqnarray}
As before we assume that the $L$ sites are occupied in the initial state. 
Thus the system contains at least one particle at later time. The  condition 
of non-emptiness can be written as
\begin{equation}
J_{m,m}(t)=0\,,\qquad m=0,L\,.
\label{e3.jmm}
\end{equation}
One may notice that according to the definitions given in~\eref{e3.ghmn},
$G_{m,m+1}(t)=H_{m,m+1}(t)=J_{m,m+1}(t)$, in agreement with 
equations~\eref{e3.jgh2} and~\eref{e3.jmm}. 

When $0<m<n<L$, the empty interval is indeed built of two disconnected parts 
and its probability satisfies the master equation
\begin{eqnarray}
\fl
{{\rm d}J_{m,n}(t)\over{\rm d}t}
&=D\,[G_{m-1,n}(t)+H_{m,n+1}(t)-G_{m,n}(t)-H_{m,n}(t)]\nonumber\\
\fl
&=D\,[J_{m-1,n}(t)-2J_{m,n}(t)+J_{m+1,n}(t)
+J_{m,n-1}(t)-2J_{m,n}(t)+J_{m,n+1}(t)]
\label{e3.mastmn}
\end{eqnarray}
where the gain terms correspond either to a particle at $m$ jumping to the 
right or a particle at $n+1$ jumping to the left and the loss terms either 
to a particle at $m+1$ jumping to the left or a particle at $n$ jumping to 
the right. When $m=0$, there is a single empty interval from site $n+1$ to 
site $L$ and the master equation reads
\begin{equation}
\fl
{{\rm d}J_{0,n}(t)\over{\rm d}t}=D\,[H_{0,n+1}(t)-H_{0,n}(t)]
=D\,[J_{0,n-1}(t)-2J_{0,n}(t)+J_{0,n+1}(t)]\,.
\label{e3.mast0n}
\end{equation}
In the same way, when $n=L$, we are left with a single empty interval from 
site $1$ to site $m$ and the corresponding probability evolves according to
\begin{equation}
\fl
{{\rm d}J_{m,L}(t)\over{\rm d}t}=D\,[G_{m-1,L}(t)-G_{m,L}(t)]
=D\,[J_{m-1,L}(t)-2J_{m,L}(t)+J_{m+1,L}(t)]\,.
\label{e3.mastml}
\end{equation}
Equations~\eref{e3.mast0n} and~\eref{e3.mastml} contain the same gain and 
loss terms as for the corresponding empty intervals in 
equation~\eref{e3.mastmn}. They remain valid for $n=L-1$ and $m=1$, 
respectively, provided $J_{m,n}(t)$ satisfies the boundary condition
\begin{equation}
J_{0,L}(t)=1\,.
\label{e3.j0l}
\end{equation}

\subsection{Solution of the eigenvalue problem}

Looking for the solutions under the form
\begin{equation}
J_{m,n}(t)=\sum_\omega\phi_\omega(m,n)\,\e^{-\omega t}\,,
\label{e3.ansatz}
\end{equation}
the master equations~(\ref{e3.mastmn}--\ref{e3.mastml}) lead to an eigenvalue 
problem which has been discussed in details in reference~\cite{krebs95}. 

Since $0\leq m<n\leq L$, there is a total of $L(L+1)/2$ modes. According 
to~\eref{e3.jmm}, $\phi_\omega(m,n)$ is an antisymmetric combination of 
eigenfunctions of the second difference operators involved 
in~(\ref{e3.mastmn}--\ref{e3.mastml}). The problem is invariant under space 
reflection so that $\phi_\omega(m,n)$ can be chosen as an eigenfunction of 
the space reflection operator
\begin{equation}
P\ :\ (m,n)\mapsto (L-n,L-m)\,.
\label{e3.p}
\end{equation}
Three types of solutions are obtained~\cite{krebs95}:
\begin{itemize}
\item The stationary solution
\begin{equation}
J_{m,n}(\infty)=\phi_0(m,n)={n-m\over L}
\label{e3.phi0}
\end{equation}
which is an eigenstate of $P$ with eigenvalue $+1$. Its expression follows 
from the fact that the zero mode eigenfunction is linear in $m$ and $n$ and 
has to satisfy the boundary conditions~\eref{e3.jmm} and~\eref{e3.j0l}. The 
first condition leads to the form $\phi_0(m,n)=c(n-m)$ and the second gives 
$c=1/L$. It has also a simple physical interpretation: since a single 
particle remains in the stationary state, a site is occupied with probability 
$1/L$. Thus the probability to have $L-n+m$ empty sites, from $1$ to $m$ and 
from $n+1$ to $L$, is given by $1-(L-n+m)/L$. 

One may notice that with $J_{0,L}(\infty)=1$, the time-dependent part of 
$J_{m,n}(t)$ has to satisfy the boundary condition
\begin{equation}
J_{0,L}(t)-J_{0,L}(\infty)=0
\label{e3.j0l0l}
\end{equation}
according to~\eref{e3.j0l}.
\item The $2(L-1)$ one-fermion excitations
\begin{eqnarray}
\fl
\phi_k^+(m,n)&={1\over\sqrt{L}}
\left[\sin\!\left({mk\pi\over L}\right)
-\sin\!\left({nk\pi\over L}\right)\right]\,,\nonumber\\
\fl
\phi_k^-(m,n)&={1\over\sqrt{L}}
\left[\left(1-{2n\over L}\right)\sin\!\left({mk\pi\over L}\right)
-\left(1-{2m\over L}\right)\sin\!\left({nk\pi\over L}\right)\right]\,,
\label{e3.phikpm}
\end{eqnarray}
with $k=1,L-1$. These functions are antisymmetric eigenstates of $P$ such 
that $P\phi_k^\pm=\pm(-1)^k\phi_k^\pm$. They vanish when $m=0$ and~$n=L$ in 
agreement with~\eref{e3.j0l0l}. The excitations energies are given by
\begin{equation}
\omega_k=4D\sin^2\!\left({k\pi\over2L}\right)\,.
\label{e3.omegak}
\end{equation}
Actually the odd eigenstate of $P$, $\phi_k^-(m,n)$, is the combination of a 
one-fermion excitation with a zero mode. 
\item The $(L-1)(L-2)/2$ two-fermion excitations
\begin{equation}
\fl
\phi_{kl}(m,n)={2\over L}
\left[\sin\!\left({mk\pi\over L}\right)\sin\!\left({nl\pi\over L}\right)
-\sin\!\left({ml\pi\over L}\right)\sin\!\left({nk\pi\over 
L}\right)\right]\,,
\label{e3.phiklmn}
\end{equation}
with $1\leq k<l\leq L-1$. These antisymmetric two-particle states are 
eigenstates of $P$ with eigenvalues $(-1)^{k+l+1}$ and they satisfy the 
boundary condition~\eref{e3.j0l0l}. The corresponding eigenvalues read
\begin{equation}
\omega_{kl}=\omega_k+\omega_l=4D
\left[\sin^2\!\left({k\pi\over2L}\right)
+\sin^2\!\left({l\pi\over2L}\right)\right]\,.
\label{omegakl}
\end{equation}
\end{itemize}
The solution satisfying the boundary conditions can be written as the 
expansion
\begin{equation}
\fl
J_{m,n}(t)\!=\!{n\!-\!m\over L}+\sum_{k=1}^{L-1}
\left[\sum_{\alpha=\pm}a_k^\alpha\,\phi_k^\alpha(m,n)\right]\,
\e^{-\omega_k t}
+\sum_{k=1}^{L-2}\sum_{l=k+1}^{L-1}\!\!b_{kl}\,\phi_{kl}(m,n)\,
\e^{-\omega_{kl}t}\,.
\label{e3.sol}
\end{equation}
All the sites are occupied with probability one in the initial state, so 
that
\begin{equation}
J_{m,n}(0)=\delta_{m,0}\,\delta_{n,L}\,.
\label{e3.condi1}
\end{equation}
Thus, for $0<m<n<L$, we have:
\begin{equation}
-\phi_0(m,n)=
\sum_{k=1}^{L-1}\sum_{\alpha=\pm}a_k^\alpha\,\phi_k^\alpha(m,n)
+\sum_{k=1}^{L-2}\sum_{l=k+1}^{L-1}\!\!b_{kl}\,\phi_{kl}(m,n)\,.
\label{e3.condi2}
\end{equation}
The coefficients of the eigenvalue expansion can be determined by making use 
of the orthogonality relations between the different eigenfunctions. 
Following reference~\cite{krebs95}, let us define surface and bulk scalar 
products for arbitrary functions $f$ and $g$ as
\begin{eqnarray}
\langle f|g\rangle_{\rm s}&=\sum_{m=1}^{L-1}f(m,L)\,g(m,L)
+\sum_{n=1}^{L-1}f(0,n)\,g(0,n)\,,\nonumber\\
\langle f|g\rangle_{\rm b}&=
\sum_{n=2}^{L-1}\sum_{m=1}^{n-1}f(m,n)\,g(m,n)\,. 
\label{e3.sp}
\end{eqnarray}
It turns out that the one-fermion eigenfunctions in~\eref{e3.phikpm} are 
orthogonal for the surface scalar product whereas the two-fermion 
eigenfunctions in~\eref{e3.phiklmn} are orthogonal for the bulk one:
\begin{eqnarray}
\langle\phi_k^\alpha|\phi_{k'}^\beta\rangle_{\rm s}
&=\delta_{kk'}\,\delta_{\alpha\beta}\qquad (k,k'=1,L-1;\ 
\alpha,\beta=\pm)\,,\nonumber\\
\langle\phi_{kl}|\phi_{k'l'}\rangle_{\rm b}
&=\delta_{kk'}\,\delta_{ll'}\qquad(0<k<l<L;\ 0<k'<l'<L)\,.
\label{e3.ortho}
\end{eqnarray}
Thus, taking appropriate scalar products in~\eref{e3.condi2}, one obtains:
\begin{eqnarray}
a_k^+&=-\langle\phi_k^+|\phi_0\rangle_{\rm s}
={2S_1(k)\over L^{3/2}}-{S_2(k)\over L^{1/2}}
=-{1+(-1)^k\over 2\sqrt{L}}\cot\!\left({k\pi\over 2L}\right)\,,\nonumber\\
a_k^-&=-\langle\phi_k^-|\phi_0\rangle_{\rm s}
={S_2(k)\over\sqrt{L}}
={1-(-1)^k\over 2\sqrt{L}}\cot\!\left({k\pi\over 2L}\right)\,,\nonumber\\
b_{kl}&=-\langle\phi_{kl}|\phi_0\rangle_{\rm b}
-\sum_{k'=1}^{L-1}\sum_{\alpha=\pm}a_{k'}^\alpha\,
\langle\phi_{kl}|\phi_{k'}^\alpha\rangle_{\rm b}
=\langle\phi_{kl}|\phi_0\rangle_{\rm b}\nonumber\\
&={(-1)^k-(-1)^l\over 2L}\cot\!\left({k\pi\over 2L}\right)
\cot\!\left({l\pi\over 2L}\right)\,.
\label{e3.coef}
\end{eqnarray}
The relation
\begin{equation}
\sum_{k'=1}^{L-1}\phi_{k'}^\pm(m,n)
\langle\phi_{k'}^\pm|\phi_0\rangle_{\rm s}=\phi_0(m,n)
\label{e3.rel}
\end{equation}
has been used in the calculation of $b_{kl}$. The final form of $J_{m,n}(t)$ 
satisfying the initial and boundary conditions follows from~\eref{e3.sol} 
and~\eref{e3.coef}.

\subsection{Crossing probability}

\begin{figure}[tbh]
\epsfxsize=7cm
\vglue3mm
\begin{center}
\mbox{\epsfbox{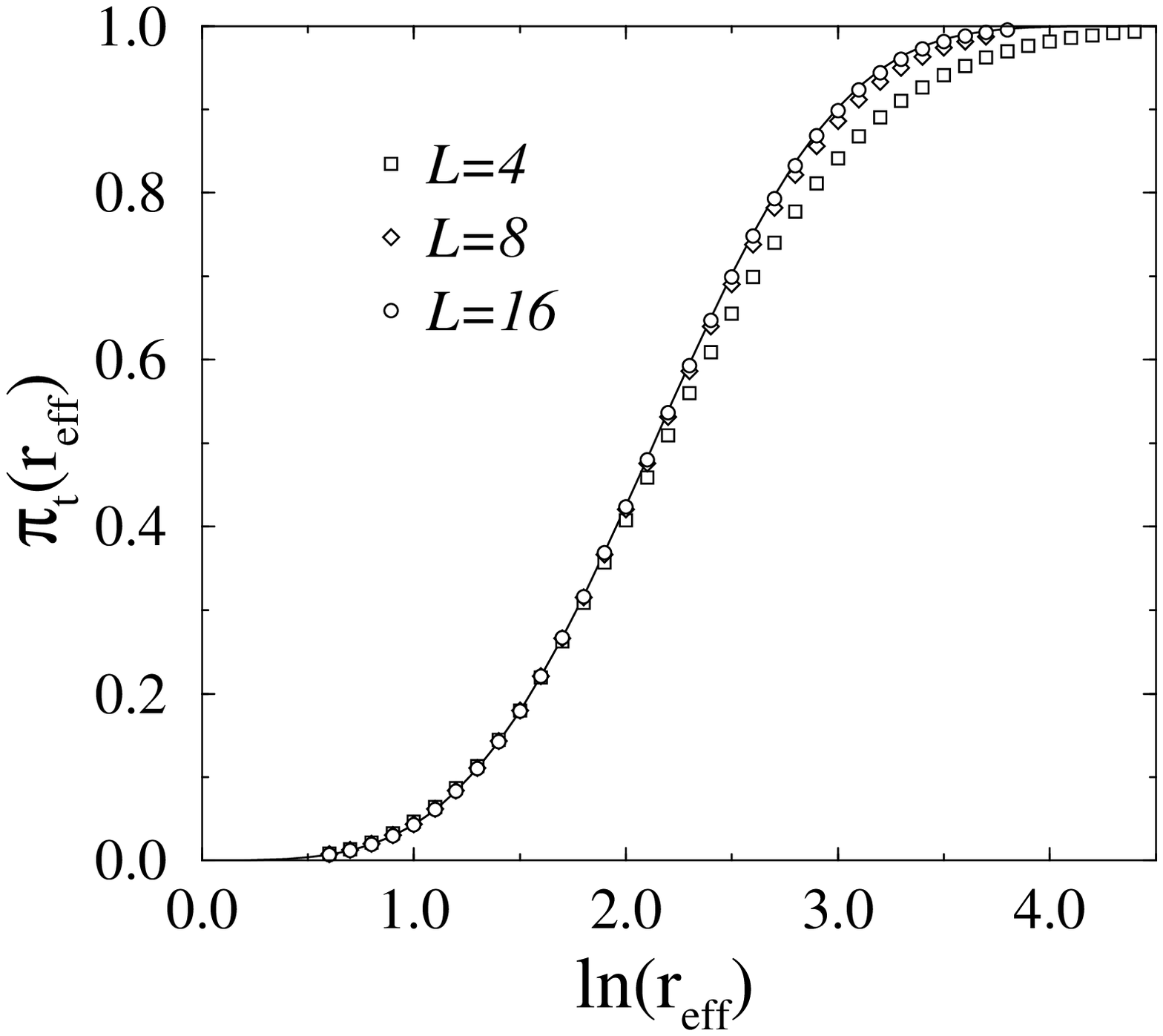}}
\end{center}
\vglue-3mm
\caption{\label{fig2} Variation of the scale-invariant crossing probability 
with the effective aspect ratio for free boundary conditions. The solid line 
corresponds to the asymptotic expression in equation~\protect\eref{e3.pit}.}
\end{figure}

Since $J_{n-1,n}(t)$ gives the probability that all the sites are empty,  
except site $n$ which is occupied by the last particle, the probability that 
the system is still active at time $t$ is given by:
\begin{eqnarray}
\fl
P_t(L,t)&\!=\!1-\sum_{n=1}^L J_{n-1,n}(t)\nonumber\\
\fl
&\!=\!-\!\sum_{k=1}^{L-1}\left[\sum_{\alpha=\pm}\!a_k^\alpha
\sum_{n=1}^L\!\phi_k^\alpha(n\!-\!1,n)\right]\!
\e^{-\omega_k t}\!
-\!\sum_{k=1}^{L-2}\sum_{l=k+1}^{L-1}\!\!b_{kl}\!
\left[\sum_{n=1}^L\!\phi_{kl}(n\!-\!1,n)\right]\!
\e^{-\omega_{kl}t}\,.
\label{e3.pilt1}
\end{eqnarray}
Straightforward but lengthy calculations lead to the following results for 
the different sums over $n$:
\begin{eqnarray}
\sum_{n=1}^L\phi_k^+(n-1,n)&=0\,,\nonumber\\
\sum_{n=1}^L\phi_k^-(n-1,n)&=-2{1-(-1)^k\over L^{3/2}}
\cot\!\left({k\pi\over 2L}\right)\,,\nonumber\\
\sum_{n=1}^L\phi_{kl}(n-1,n)&={1-(-1)^{k+l}\over L}
{\sin\!\left({k\pi\over L}\right)\sin\!\left({l\pi\over L}\right)\over
\sin\!\left[{(k-l)\pi\over 2L}\right]
\sin\!\left[{(k+l)\pi\over 2L}\right]}\,.
\label{e3.sums}
\end{eqnarray}
Thus we obtain 
\begin{eqnarray}
\fl
P_t(L,t)&\!=\!{4\over L^2}
\sum_{\scriptstyle k=1\atop\scriptstyle k\;\mbox{\scriptsize odd}}^{L-1}
\cot^2\!\left({k\pi\over 2L}\right)\cot^2\!\left({l\pi\over 2L}\right)
\exp\!\left[-4Dt\sin^2\!\left({k\pi\over 2L}\right)\right]\nonumber\\
\fl
&-{8\over L^2}\!\sum_{k=1}^{L-2}\!\!
\sum_{\scriptstyle l=k+1\atop\scriptstyle k+l\;
\mbox{\scriptsize odd}}^{L-1}\!\!\!(\!-1)^k
{\cos^2\!\left({k\pi\over 2L}\right)\cos^2\!\left({l\pi\over 2L}\right)\over
\sin\!\!\left[{(k-l)\pi\over 2L}\right]
\!\sin\!\!\left[{(k+l)\pi\over 2L}\right]}
\exp\!\left\{\!\!-4Dt\!\left[\sin^2\!\left(\!{k\pi\over 2L}\!\right)
\!\!+\!\sin^2\!\left(\!{l\pi\over 2L}\!\right)\!\right]\!\right\}\!\!.
\label{e3.pilt2}
\end{eqnarray}
In the finite-size scaling limit, the crossing probability displays the 
scale-invariant dependence on the effective aspect ratio $r_{\rm eff}=L^2/Dt$ 
and reads
\begin{eqnarray}
\fl
\pi_t(r_{\rm eff})\!&=\!\lim_{\scriptstyle L,t\to\infty\atop
\scriptstyle r\;\mbox{\scriptsize fixed}}P_t(L,t)\nonumber\\
\fl
&=\!{16\over\pi^2}\!\!
\sum_{\scriptstyle k=1\atop\scriptstyle k\;\mbox{\scriptsize odd}}^\infty
\!\!{1\over k^2}\exp\!\left(\!-{k^2\pi^2\over r_{\rm eff}}\right)
\!-\!{32\over\pi^2}\sum_{k=1}^\infty\!\!
\sum_{\scriptstyle l=k+1\atop\scriptstyle k+l\;
\mbox{\scriptsize odd}}^\infty\!\!{(-1)^k\over k^2-l^2}
\exp\!\left[-{(k^2\!+\!l^2)\pi^2\over r_{\rm eff}}\right]\!\!+\!\Or(L^{-2}).
\label{e3.pit}
\end{eqnarray}
The rapid convergence to the scaling limit is shown in figure~\ref{fig2}.

%%%%%%%%%%%%%%%%%%%%%%%%%%%%%%%%%%%%%%%%%%%%%%%%%%%%%%%%%%%%%%%%%%%%
%%%%%%%%%%%%%%%%%%%%%%%%%%%%%%%%%%%%%%%%%%%%%%%%%%%%%%%%%%%%%%%%%%%%

\section{Discussion}

We have studied DLC at the critical point where the particle density and 
other quantities display power laws in the thermodynamic limit. The problem 
can be made off-critical by introducing a birth process with rate $\Delta$, 
corresponding to the back-reaction of the coagulation process 
in~\eref{e2.proc}. The system being strongly anisotropic, under a change of 
the length scale by a factor $b$, the length transforms as $L'=L/b$ whereas 
the time transformation, $t'=t/b^z$, involves the anisotropy or dynamical 
exponent $z$~\cite{binder89}. For a scale-invariant crossing probability one 
obtains
\begin{equation}
P_t(L,t,\Delta)
=P_t\!\left({L\over b},{t\over b^z},b^{1/\nu}\Delta\right)\,,
\label{e4.scal1}
\end{equation}
where $\nu$ is the exponent of the correlation length, 
$\xi=\hat{\xi}\,\Delta^{-\nu}$, which diverges at the critical point, 
$\Delta=0$. Taking $b=\Delta^{-\nu}$ leads to
\begin{equation}
P_t(L,t,\Delta)=P_t\!\left({L\over\Delta^{-\nu}},{t\over\Delta^{-z\nu}},1
\right)=f\!\left({L\over\xi},{t\over\tau}\right)\,,
\label{e4.scal2}
\end{equation}
where we introduced the relaxation time, $\tau=\hat{\tau}\,\Delta^{-z\nu}$. 
Both the correlation length and the relaxation time contain a non-universal 
amplitude, $\hat{\xi}$ and $\hat{\tau}$, respectively. In 
equation~\eref{e4.scal2}, $f(x,y)$ is a universal scaling function of its 
dimensionless variables, $x=L/\xi$ and $y=t/\tau$. The finite-size scaling 
limit at criticality amounts to take $\Delta=0$ and $b=L$ in~\eref{e4.scal1}, 
which gives
\begin{equation}
P_t(L,t,0)=P_t\!\left(1,{t\over L^z},0\right)=\pi_t\!\left(c\,{L^z\over 
t}\right)\,.
\label{e4.scal3}
\end{equation}
Thus the crossing probability $\pi_t$ is a scale-invariant universal 
function of the effective aspect ratio $c\,r$. The non-universal amplitude 
$c$ depends on the choice of the length and time units. It can be expressed 
as a function of the non-universal correlation length and relaxation time 
amplitudes~\cite{hucht02} by comparing~\eref{e4.scal3} to~\eref{e4.scal2}. 
Since $r$ appears through the dimensionless ratio $x^z/y$, one obtains
\begin{equation}
c\,r={(L/\xi)^z\over t/\tau}={\hat{\tau}\over\hat{\xi}^z}\,r
\label{e4.cr}
\end{equation}
and $c=\hat{\tau}/\hat{\xi}^z$ which is equal to $D^{-1}$ in our case.

We have shown on the example of DLC that the  crossing probability $\pi_t$ 
is a scale-invariant function of the effective aspect ratio for different 
types of boundary conditions. This function is expected it to be universal as 
in the case of directed percolation~\cite{turban02}. An indication of the 
universality of $\pi_t$ can be found in~\cite{henkel01} where a birth process 
of the form $A\,\varnothing\,A\ \stackrel{2\lambda D}{\longrightarrow}\ 
A\,A\,A$ was added to~\eref{e2.proc}. The problem stays in the same 
universality class and remains exactly solvable through the empty interval 
method for all values of $\lambda$. It turns out that $I_n(t)$ is only 
modified by terms of higher order in $1/L$ which disappear in the finite-size 
scaling limit. One could also check the universality of $\pi_t$ on the 
diffusion-annihilation problem, which has been shown to belong to the same 
universality class as DLC, through a similarity transformation in the quantum 
Hamiltonian formulation of the master equation~[38--40]. 
Another possibility would be to verify the independence on the initial 
conditions, provided a finite fraction of the sites are occupied in the 
initial state.

Finally, let us mention that a recent generalization of local scale 
invariance for strongly anisotropic systems~\cite{pleimling01,henkel02} 
leaves the hope for directly obtaining the crossing probability formulas in a 
given universality class, like in isotropic systems.    

\ack

The author wishes to thank Malte Henkel for valuable discussions. 
The Laboratoire de Physique des Mat\'eriaux is Unit\'e Mixte de 
Recherche CNRS No~7556.

\end{document}